\documentstyle[psfig]{aipproc}
\begin{document}

\title{Atmospheric Neutrinos from Charm}
\author{L. Pasquali$^1$, M. H. Reno$^{1}$ and I. Sarcevic$^2$}
\address{
$^1$Department of Physics and Astronomy, University of Iowa, Iowa City,
Iowa 52242\\
$^2$Department of Physics, University of Arizona, Tucson, Arizona
85721}

\maketitle

\begin{abstract}
Using next-to-leading order perturbative QCD, we find that the charm 
contribution to the atmospheric lepton fluxes dominates over the conventional 
ones from $\pi$ and $K$ decays for energies higher than 10$^5$ GeV. 
We also discuss 
theoretical uncertainties involved in the calculations and compare our 
results with previous evaluations.

\end{abstract}

\section*{INTRODUCTION}

Conventional lepton fluxes from $\pi$ and $K$ decays 
in the atmosphere are well known \cite{gaisser}, 
but at energies of a few GeV charm-anticharm pairs can be produced and the 
leptons coming from charm decays, called prompt leptons, start to contribute 
to the atmospheric lepton fluxes. 
Results from several experiments \cite{frejus}
show a muon excess relative to the conventional 
muon flux in the 10 TeV energy range that could be an indication of a charm 
contribution. Large volume water and ice neutrinos experiments \cite{amanda}
are designed to search for muon neutrinos from extragalactic sources for 
which atmospheric muons and muon neutrinos represent the main background. 
Therefore it is important to determine the atmospheric lepton flux,
including charm contributions, with 
better accuracy.

Atmospheric lepton fluxes from charm decays have been evaluated in the 
past using different models of charm production \cite{8,others}. We 
present a new calculation \cite{prs}
based on next-to-leading order (NLO) perturbative QCD. 
We also investigate the importance of the small-$x$ behaviour of the parton 
distribution functions and study the uncertainties involved in the necessary 
extrapolation of cross sections and energy distributions beyond the 
experimentally measured regime.

\section*{Lepton Flux Calculation}

The lepton fluxes can be calculated semi-analytically using approximate
cascade equations which begin with the incident cosmic ray flux. 
Details of this solution method can be found, for 
example, in Refs. \cite{lip,prs}. 
As in  Ref. \cite{8}, we assume that the incident cosmic ray flux can 
be represented by protons and that at the top of the atmosphere ({\it X}=0) 
the proton flux $\phi_p(E,X)$ scales as $1.7\ (E/{\rm GeV})^{-2.7}$ 
for $E<5\cdot 10^6$ GeV and scales as $174\ (E/{\rm GeV})^{-3}$ for 
$E\ge 5\cdot 10^6$ GeV.

Production of charmed hadrons $j$ of energy $E$ by proton-air interactions
enters into the calculation via the ``$Z$-moment'',
expressed in terms of an integral over 
$x_E=E/E_p$, which is given by
\begin{equation}
Z_{pj}=2f_j\int_0^1{dx_E\over x_E}{\phi_p(E/x_E)\over\phi_p(E)}{1\over
\sigma_{pA}(E)}{d\sigma_{pA\rightarrow c\bar{c}}(E/x_E)\over
dx_E}, 
\end{equation}
where $f_j$ is the fraction of charmed particles which emerge as hadron $j$ 
and the factor of two is taking into account the fact that we implicitly 
sum over particles and antiparticles. The quantity $\sigma_{pA}(E)$ is the
total proton-air cross section \cite{mielke}.

The charm production cross section and energy distribution depends
on the value of the charm mass ($m_c$), the
factorization scale ($M$) and renormalization scale ($\mu$) 
as well as the choice of parton distribution functions.
Comparing the NLO total charm cross section \cite{nlo} with the 
experimental data from Ref. \cite {13} led us to set
$m_c = 1.3$ GeV. We use the CTEQ3 \cite{15} 
and the MRSD- \cite{16} parton distribution functions. 
We set our default factorization and renormalization scales to
$M=2\mu=2m_c$.
We also evaluate the lepton flux using the CTEQ3 parton distribution 
functions with $M = \mu = m_c$. 

Because of the large computing time involved, we have used
the NLO Monte Carlo program from Mangano, Nason and 
Ridolfi \cite{14} to evaluate the appropriate rescaling factor $K$ that, 
applied to the ``LO'' energy distribution, correctly accounts for the NLO 
corrections. Here ``LO'' means taking the leading order matrix element squared,
but using the two-loop $\alpha_s(\mu^2)$ and the NLO parton distribution
functions. The parametrization of $K(E,x_E)$ is given by the 
following expression
\begin{eqnarray}
K(E,x_E) & = & 1.36+0.42\ln(\ln(E/{\rm GeV}))\\ \nonumber
& &\quad +\Bigl( 3.40+
18.7(E/{\rm GeV})^{-0.43}-0.079\ln(E/{\rm GeV})\Bigr)\cdot x_E^{1.5}.
\end{eqnarray}
The $K$-factor rescaled ``LO'' charm production distribution is used
in Eq. (1) to obtain our lepton fluxes.

In Fig. 1 we show our results for the prompt atmospheric muon flux
scaled by $E^3$ for two parton distributions and factorization
scale choices.  The highest flux  at
$E=10^8$ GeV is with the MRSD- distribution and
$M=2\mu=2 m_c$ (dashed). The CTEQ3 distributions with the same choice
of scale are represented by the solid line, while the dot-dashed line
shows the result when $M=\mu=m_c$. For reference, we show the vertical
conventional and prompt muon flux calculated and parameterized by Thunman
{\it et al.} in Ref. \cite{8}. The prompt flux is isotropic.
The prompt lepton flux evaluated using perturbative QCD can be
parameterized as 
\begin{equation}
\log_{10}\Bigl( E^3\phi_\ell (E)/({\rm GeV}^2/
{\rm cm}^2\,{\rm s\, sr})\Bigr)
 = -A + B\, x+C\,x^2-D\, x^3,
\end{equation}
where $x\equiv \log_{10}(E/{\rm GeV})$.
In Table I, we collect the constants for the MRSD- and CTEQ3 fluxes
exhibited in Fig. 1. The prompt neutrino fluxes equal the prompt muon flux.

% TABLE

\begin{table}
\caption{Parameters for the prompt
muon plus antimuon flux appearing in Fig. 1, parameterized by
Eq. (3).}
\begin{tabular}{llcccc}

PDF & Scales & $A$ & $B$ & $C$ & $D$ \\ \hline
CTEQ3 & $M=\mu=m_c$ & 5.37 & 0.0191 & 0.156 & 0.0153 \\
CTEQ3 & $M=2\mu=2m_c$ & 5.79 & 0.345 & 0.105 & 0.0127 \\
MRSD- & $M=2\mu=2m_c$ & 5.91 & 0.290 & 0.143 & 0.0147 \\
\end{tabular}
\end{table}

\section*{Conclusions}
\begin{figure}
\centerline{\psfig{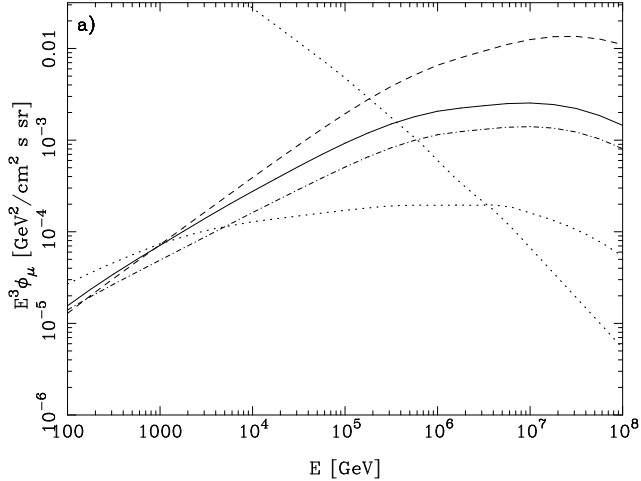}}
\caption{The prompt atmosheric muon flux scaled by $E^3$ versus muon energy 
for CTEQ3 (solid) and MRSD- (dashed) with $M = 2\mu = 2m_c$. 
Also shown is the scaled muon flux using CTEQ3 with $M = \mu = m_c$ 
(dot-dashed) and the Thunman {\it et al.} parametrization [4] of the prompt 
muon flux and the vertical conventional muon 
flux (dotted).
}
\end{figure}
We find that the perturbative charm contributions to lepton
fluxes are significantly larger than the recent Thunman
{\it et al.}  calculation \cite{8}.
The prompt muon flux becomes larger than the conventional muon
flux from pion and kaon decays at energies above $\sim 10^5$ GeV. We have evaluated the energy distribution at NLO in QCD, and
we find that the NLO corrections give a correction
of more than a factor of two which is weakly energy and $x_E$
dependent. 

The main uncertainty in the perturbative calculation of the
prompt flux, given fixed charm mass, factorization scale and
renormalization scale, is the small-$x$ behavior of the parton
distribution functions. The spread in predictions at $E=10^8$ GeV
is indicative of this uncertainty.

We conclude that the prompt muon flux calculated in the context
of perturbative QCD cannot explain the observed excess of
muons in the TeV region \cite{frejus}, independent
of the theoretical uncertainties associated with small
parton $x$. Measurements of the atmospheric flux in the 100 TeV
range would help pin down the charm cross section at energies
above those currently accessible using accelerators and would
provide valuable information about the small-$x$ behavior of
the gluon distribution function.

\section*{Acknowledgements}
Work supported in part
by National Science Foundation Grant Nos.
PHY-9507688 and PHY-9802403 and D.O.E. Contract No. 
DE-FG02-95ER40906.

\end{document}